\documentstyle[aas2pp4]{article}

\def\ni{\noindent}

\def\K{{\rm\,K}}
\def\km{{\rm\,km}}

\def\cm{{\rm\,cm}}

\def\yr{{\rm\,yr}}
\def\Gyr{{\rm\,Gyr}}

\def\g{{\rm\,g}}

\def\s{{\rm\,s}}
\def\ms{{\rm\,ms}}

\def\ergs{{\rm\,ergs}}

\def\keV{{\rm\,keV}}
\def\MeV{{\rm\,MeV}}
 
\def\G{{\rm\,G}}

\begin{document}

\title{Constraining dense-matter superfluidity through thermal emission
from millisecond pulsars}

\author{Andreas Reisenegger}

\affil{Departamento de Astronom\'\i a y Astrof\'\i sica, 
Facultad de F\'\i sica,\\
Pontificia Universidad Cat\'olica de Chile,\\
Casilla 104, Santiago 22, Chile}

\authoremail{areisene@astro.puc.cl}

\begin{abstract}
As a neutron star spins down, the gradual decrease of the centrifugal force 
produces a progressive increase of the density of any given fluid element
in its interior. Since the ``chemical'' (or ``beta'') equilibrium state is 
determined by the local density, this process leads to a chemical imbalance 
quantified by a chemical potential difference, e.g.,  
$\delta\mu\equiv\mu_n-\mu_p-\mu_e$, where $n$, $p$, and $e$ denote neutrons,
protons, and electrons. In the presence
of superfluid energy gaps, in this case $\Delta_n$ and $\Delta_p$, reactions
are strongly inhibited as long as both $\delta\mu$ and $kT$ are much smaller
than the gaps. Thus, no restoring mechanism is available, and the imbalance 
will grow unimpeded until $\delta\mu=\delta\mu_{thr}\sim\Delta_n+\Delta_p$. 
At this threshold, the reaction rate increases dramatically, preventing further
growth of $\delta\mu$, and converting the excess chemical energy into heat.
The thermal luminosity resulting from this ``rotochemical heating'' process is 
$L\sim 2\times 10^{-4}(\delta\mu_{thr}/0.1\MeV)\dot E_{rot}$, similar to the typical
x-ray luminosity of pulsars with spin-down power $\dot E_{rot}$. The threshold
imbalance, and therefore the luminous stage, are
only reached by stars whose initial rotation period is 
$P_i\lesssim 12(\delta\mu_{thr}/0.1\MeV)^{1/2}\ms$, i.e., millisecond pulsars.
A preliminary study of eleven millisecond pulsars with reported ROSAT observations 
shows that the latter can already be used to start constraining
superfluid energy gaps in the theoretically interesting range, $\sim 0.1 - 1
\MeV$.

\end{abstract}


\section{Introduction}

The precise timing of radio pulsars allows an excellent determination of 
neutron star 
rotation periods, $P$, and their (positive) time derivatives, $\dot P$, which in 
turn yield the characteristic time scales for spin-down, 
$t_s\equiv P/(2\dot P)\sim 10^3 - 10^{10}\yr$ (Taylor, Manchester, \& Lyne 1993). 
At least some of these objects are born in supernova explosions, after which 
they are hot ($T\sim 10^{11}\K$) and spin with periods of a few
tens of milliseconds. Standard models of cooling through neutrino and photon
emission leave neutron stars with negligible internal heat after 
$\sim 10^7\yr$ (Nomoto \& Tsuruta 1987; Page \& Applegate 1992), while they 
gradually spin down. 

{\it Millisecond pulsars} are faster 
($P\sim 1.5 - 10\ms$), older ($t_s > 10^8\yr$), and less magnetized (with 
surface magnetic field strengths $B\sim 10^{8-9}\G$) than ordinary pulsars, 
and usually have binary companions, most often cool white dwarfs (Phinney 
\& Kulkarni 1994). This combination of properties has led to the suggestion
that these stars have been ``recycled'' through accretion from a binary 
companion, which gave them additional angular momentum and, perhaps,
reduced their magnetic field (Bisnovatyi-Kogan \& Komberg 1975; Bhattacharya
\& van den Heuvel 1991). The temperatures of the white dwarf companions, when 
compared with cooling models, provide 
an independent age estimate, $t_c$, which tends to be smaller than $t_s$, 
though not by a large factor (Kulkarni 1986; Bell, Bailes, \& Bessell 1993; 
Lorimer et al. 1995; Bell et al. 1995; Alberts et al. 1996; Lundgren et al. 
1996a; Lundgren, Foster, \& Camilo 1996b). This confirms 
that these are old systems, in which {\it the pulsar has already lost a 
considerable amount of rotational energy} (similar or somewhat smaller than 
that which it currently carries) {\it and 
essentially all the thermal energy it originally contained.}

Neutron star cores are usually assumed to contain neutrons,
protons, electrons, in the denser central region also muons, and possibly
other kinds of particles. More exotic possibilities, such as
``free'' quark cores or even ``strange stars'' composed exclusively
of a mixture of $u$, $d$, and $s$ quarks not confined into hadrons
(plus a small admixture of electrons) also cannot be excluded at present.

Within the standard scenario, it was proposed decades ago that both neutrons 
and protons should form Cooper pairs (bound by strong interactions) and 
undergo a {\it superfluid transition} at a still very uncertain temperature 
$T_c\sim 10^{9-10}\K$ (Migdal 1960; Ginzburg \& Kirzhnits 1965; Sauls 
1989), much lower than the nucleon Fermi temperatures $T_F\sim 10^{12}\K$, 
but higher than temperatures expected inside observed neutron stars 
($T\lesssim 10^8\K$), and particularly in ms pulsars. In spite of the
long and distinguished history of interest in the subject, there has been
little observational evidence for it, mostly limited to glitch dynamics,
which is interpreted in terms of pinning and unpinning of neutron superfluid 
vortices to the solid lattice in the inner crust (Pines \& Alpar 1985).

In the superfluid state, the density of
quasiparticle states as a function of energy has a {\it gap} 
between $E_F-\Delta$ and $E_F+\Delta$, where $E_F$ is the Fermi energy and
$\Delta$ is the gap parameter (Tilley \& Tilley 1990). Neutrons 
in the stellar 
crust and protons in the core are expected to have an isotropic gap 
(with $\Delta\approx 1.76 kT_c$ at $T=0$) like that in laboratory 
superconductors, whereas core neutrons should have a more complicated, 
anisotropic gap (Amundsen \& \O stgaard 1985). At low temperatures, 
$kT\ll\Delta$, the number of occupied states above the gap and that 
of empty states below it are proportional to $\sim \exp(-\Delta/kT)$, 
which suppresses the heat capacity and the equilibrium reaction rates by 
similar factors (Levenfish \& Yakovlev 1992a,b),
strongly affecting the thermal evolution of neutron stars. There have
been attempts to use x-ray observations of cooling neutron stars to 
constrain the energy gaps (Page \& Applegate 1992; Page 1994, 1995), but so 
far the results are not very conclusive (Page 1996).

In this paper, I discuss how the presence of superfluid energy gaps delays
and enhances the recently proposed process of {\it ``rotochemical heating''}
(Reisenegger 1995; hereafter R95), leading to {\it substantial thermal emission
from ancient millisecond pulsars}. The presence or absence of this emission can
therefore be used to constrain the value of the gaps.
The next section describes the physical model, which is applied to observational
data in \S 3 to obtain preliminary constraints. In \S 4 the results are
discussed and improvements are suggested.

\section{The model}
 
The fast rotation rate, $\Omega=2\pi/P$, of ms pulsars produces a strong 
centrifugal force, which reduces their internal density by 
$\Delta\rho/\rho\approx -\alpha\Omega^2/(2G\rho_c)$ with respect to its 
nonrotating value. (Here, $\rho_c$ is the central density, $G$ is the 
constant of gravity, and $\alpha$ is a dimensionless
number of order unity [R95]). {\it Their spin-down causes a slow 
contraction.} The relative abundances of different particle species
in reaction equilibrium in the stellar core depend on the density,
and therefore {\it the contraction gives rise to a ``chemical'' imbalance}, 
whose effect on the thermal evolution has been considered 
for the cores of neutron stars (R95) and strange stars (Cheng \& Dai 1996),
as well as for neutron star crusts (Iida \& Sato 1996), in all cases
without superfluidity. In most of this section, I concentrate on neutron 
star cores of standard
composition (neutrons, protons, and electrons), in which direct Urca 
reactions ($n\rightarrow p+e+\bar\nu_e$ and $p+e\rightarrow n+\nu_e$)
are allowed (Lattimer et al. 1991), and in which {\it the neutrons,
the protons, or both are in the superfluid state}, with gap parameters
$\Delta_n$ and $\Delta_p$. Alternatives are discussed below.

The chemical imbalance can be quantified by $\delta\mu=\mu_n-\mu_p-\mu_e$, 
where $\mu_i$ are the chemical potentials ($\approx$ Fermi energies) of 
the three particle species 
($i=n,p,e$; note the opposite sign convention than in R95).
In non-superfluid matter, the diffusion of charged particles with respect
to the neutrons is limited by proton-neutron collisions, with mean free
time $\tau_{pn}\sim 3\times 10^{-17}T_8^{-2}\s$ (Yakovlev \& Shalybkov 1990), 
where $T_8\equiv T/10^8\K$. The time scale for protons moving at their Fermi
velocity, $v_{Fp}\sim 0.2 c$, to diffuse a neutron star radius, $R\sim 10\km$,
is $t_{pd}\sim (R/v_{Fp})^2\tau_{pn}^{-1}\sim 30 T_8^2 \yr$, much shorter 
than the evolutionary time scale of a ms pulsar, $t_s\gtrsim 10^8\yr$.
This inequality is likely to become even stronger (or at least not much weaker)
if superfluid matter is considered instead. Thus, the stellar core
can be regarded as being in {\it diffusive equilibrium}, with $\delta\mu$
constant throughout the stellar core.

\begin{figure}
\plotone{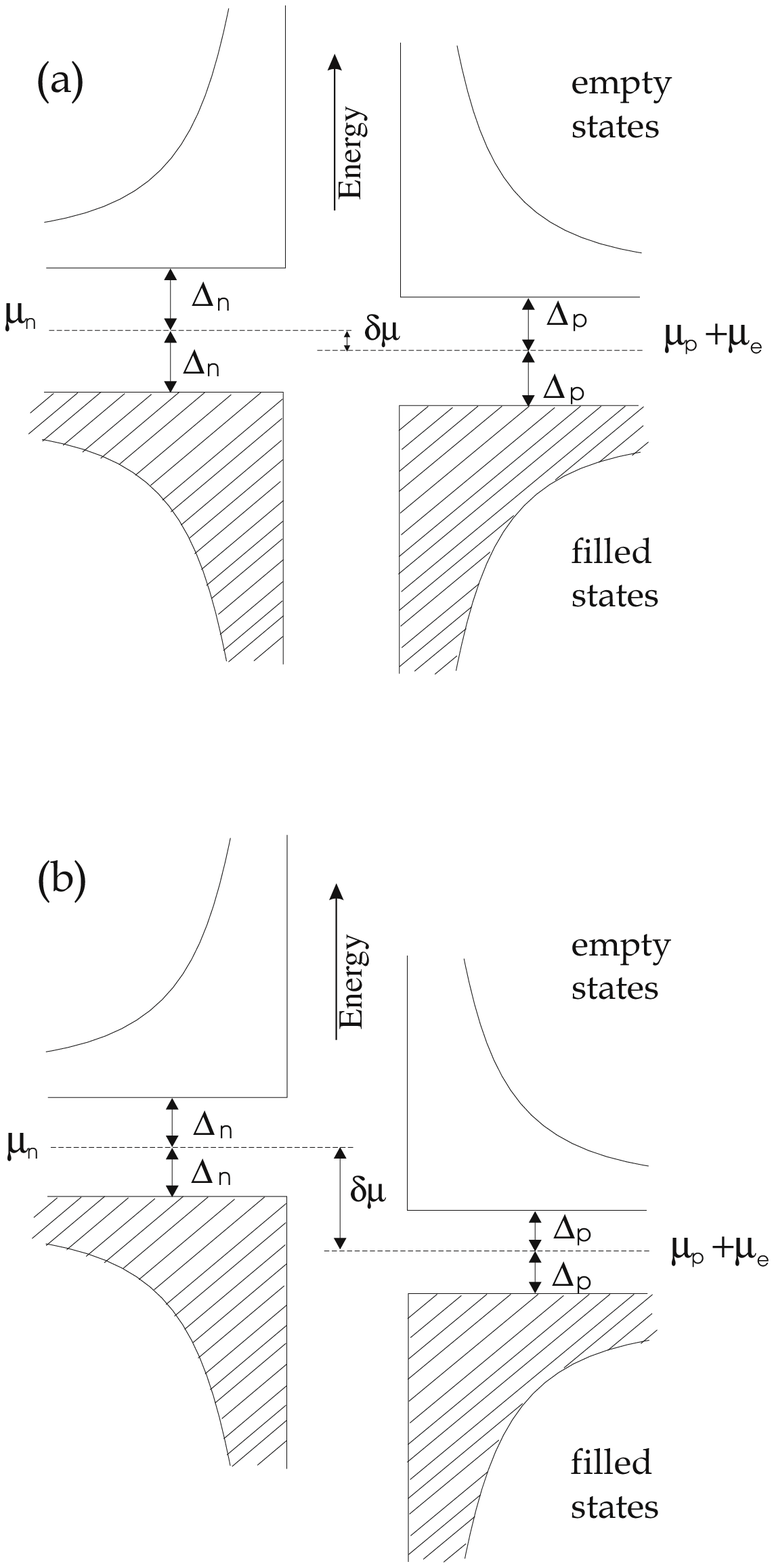}
\caption{Schematic representation of the condition for the occurrence
of neutron beta decays (direct Urca). 
(a) $\delta\mu<\delta\mu_{thr}=\Delta_n+\Delta_p$: Reactions not allowed. 
(b) $\delta\mu>\delta\mu_{thr}=\Delta_n+\Delta_p$: Reactions allowed.}
\end{figure}

The reactions are suppressed by the superfluid energy gap(s), so
$|\delta\mu|$ can at first freely grow as
$\dot{\delta\mu}=-\alpha(q-1/3)\mu_e\Omega\dot\Omega/(G\rho_c)$,
where $q$ is another parameter of order unity, defined as the logarithmic 
derivative of the symmetry energy with respect to density 
(see R95 for details). For definiteness, I will assume
in the following discussion that $q>1/3$, so that the contraction 
produces a neutron excess and a positive $\delta\mu$. As shown schematically
in Figure 1, substantial reactions are only ``turned on'' once 
$\mu_n-\Delta_n\approx\mu_p+\Delta_p+\mu_e$,
i.e. when a neutron just below the neutron energy gap is energetically
allowed to decay into a proton just above the proton energy gap, an electron 
close to the Fermi sphere, and an antineutrino of arbitrarily small energy. 
Once this threshold, $\delta\mu_{thr}\approx\Delta_n+\Delta_p$, is exceeded 
in some region of the star
(and for some spatial direction, if one or both of the gaps are anisotropic), 
the reaction rate in that region increases rapidly, preventing 
further growth of $\delta\mu$. The threshold value is reached once the 
star has lost a rotational energy 
$\Delta E_{rot}\approx 1.4\times 10^{50}(\delta\mu_{thr}/0.1\MeV)\ergs$
(assuming a moment of inertia $I=10^{45}\g\cm^2$), i.e., 
the total rotational energy of a star rotating 
with $P\sim 12(\delta\mu_{thr}/0.1\MeV)^{-1/2}\ms$. Therefore, only ms 
pulsars are likely to ever reach this stage. 

At all later times, the reaction rate
is very nearly as needed to keep the imbalance at threshold, 
$\delta\mu=\delta\mu_{thr}$, and each reaction leaves a thermal energy 
$\approx\delta\mu_{thr}$ in the stellar core, with a much smaller amount
escaping as neutrinos. The thermal energy must eventually
reach the surface of the star and be radiated as thermal
photons, with a bolometric luminosity

\begin{equation}
L\sim \alpha(3q-1)xN\delta\mu_{thr}\,{1-2x\over 1+4x}\,
{\Omega\dot\Omega\over G\rho_c},
\end{equation}

\ni where $N$ is the total number of baryons in the star, and $x$ is the proton
fraction ($x>1/9$ for direct Urca reactions to be allowed). 
$L$ is conveniently expressed as a fraction
of the total spin-down power, i.e., the rotational energy loss per unit time,
$\dot E_{rot}=I\Omega\dot\Omega$.
For typical parameter values (see R95), 
$L/\dot E_{rot}\sim 2\times 10^{-4}(\delta\mu_{thr}/0.1\MeV)$. 

If $x<1/9$, so that direct Urca reactions are forbidden by simultaneous 
energy and momentum conservation, the dominant reactions are modified
Urca reactions, in which an additional particle participates (via strong
interactions), carrying off the excess momentum (Chiu \& Salpeter 1964). As 
long as no other hadrons are present, this extra particle will be a neutron or
a proton, whichever has the smaller energy gap, $\Delta_<$, and the threshold 
chemical imbalance becomes $\delta\mu_{thr}\approx\Delta_n+\Delta_p+2\Delta_<$.
Otherwise, the discussion given for direct Urca processes goes through 
unchanged. For modified Urca processes 
mediated by other particles, such as kaons or pions (where $\delta\mu_{thr}$
should be as in the direct Urca case), and in quark matter (where
superfluid quarks may take the place of the superfluid nucleons),
the effect should also be similar, but with a somewhat later onset,
due to the more homogeneous composition of the stellar core.

For comparison, the steady-state luminosity of non-superfluid neutron stars
is given by $L/\dot E_{rot}\sim 4\times 10^{-5}
(\dot E_{rot}/3\times 10^{33}\ergs\s^{-1})^{1/7}\MeV$ 
if only modified Urca reactions are allowed, and much lower if fast cooling
processes are present (R95). Thus, superfluid and non-superfluid neutron stars
should be easily distinguishable by this effect.

The whole discussion above is based on the assumption that the particle 
species in the stellar
core can only achieve chemical equilibrium by reactions occuring inside the
core. However, as said above, the particles can easily move throughout the star
on time scales much shorter than the evolutionary times of ms pulsars. Thus,
parts of any neutron excess in the core can easily flow into the crust,
where they can be absorbed by nuclei. This process is opposite to the 
neutron emission and absorption by nuclei in the inner crust as its
density changes due to the spin-down process (Iida \& Sato 1996). However,
since many more neutrons are available in the core than in the crust, the
former are likely to dominate and completely change the nuclear chemistry
in the inner crust. Depending on the time scales and thresholds for the
relevant nuclear transformations in the crust, this process may provide a
shortcut for the chemical equilibration and prevent the imbalance from growing
as much as assumed in the previous sections and in R95. This is clearly a 
subject that deserves more study.

\section{Application to observational data}

In this section, I make a preliminary analysis of the data available in the
literature for eleven millisecond pulsars observed (but in most cases not detected) 
in the soft x-rays by ROSAT (Kulkarni et al. 1992; Fruchter et al. 1992;
Becker \& Tr\"umper 1993; Danner, Kulkarni, \& Thorsett 1994; Halpern, Martin,
\& Marshall 1996; Halpern 1996; Becker et al. 1996; Verbunt et al. 
1996)\footnote{I omit PSR B1620-21 from the analysis
since no intrinsic period derivative, $\dot P$, is available for it.}, in 
order to show how the threshold chemical
imbalance, $\delta\mu_{thr}$, can be constrained by observations. Given
the limitations of the present study (discussed in detail below), any given 
limit determined here should not be taken too seriously, but their combination
gives an indication of what could be done.

For pulsars that have {\it not} yet reached $\delta\mu_{thr}$, the chemical 
disequilibrium at the present time can be approximated by

\begin{equation}
\delta\mu\approx\delta\mu_1\equiv 18\left(P\over\ms\right)^{-2}
{t\over t_s-t}\MeV\lesssim\delta\mu_{thr},
\end{equation}

\ni where $t$ is the true age of the pulsar (more precisely, the time since
it left the chemical equilibrium state), and magnetic dipole braking
($\dot\Omega\propto -\Omega^3$) has been assumed. Alternatively, if
the threshold value {\it has} been reached,

\begin{equation}
\delta\mu=\delta\mu_{thr}\lesssim\delta\mu_2\equiv 500{L\over\dot E_{rot}}\MeV,
\end{equation}

\ni where $L$ is the bolometric luminosity of the pulsar, and the inequality
allows for additional heating or other emission processes. For a given
pulsar, it cannot be assessed a priori which of these applies, so it
can only be asserted that {\it either} $\delta\mu_{thr}\gtrsim\delta\mu_1$
{\it or} $\delta\mu_{thr}\lesssim\delta\mu_2$.

The parameters needed to estimate $\delta\mu_2$ are listed in Table 1.
Since all of these pulsars are heavily absorbed at the low-energy end of 
the ROSAT energy band, $0.1\keV$ (which is of most interest
for the present application, since the effective temperatures turn out to be
quite low), it is important to have a good estimate of the column density
of neutral hydrogen along the line of sight to the pulsar, $N_{HI}$, in order
to infer the true luminosity from the observed counts. 

The first group of (seven) pulsars is at medium to high Galactic latitudes, 
$|b|>15^\circ$, therefore
a reasonable estimate of $N_{HI}$ is given by the velocity-integrated 
21 cm emission near the position of the pulsars, which yields the
values given in Table 1. For all of these pulsars, this value is within
a factor of 4 (and in three cases less than a factor of 1.15) of the 
``rule of thumb'' $N_{HI}=10 N_e$, where $N_e$ is the electron column density 
inferred from the pulsar dispersion measure. 

%The three more distant pulsars in this sample are within $15\%$ of the rule of 
%thumb $N_{HI}=10 N_e$, where $N_e$ is the electron column density inferred
%from the pulsar dispersion measure. This is not true for J0437-4715
%($N_{HI}\approx 24 N_e$), which is so close that it is likely to be affected 
%by local variations in the densities of the two components, and for 
%J1012+5307 ($N_{HI}\approx 2.8 N_e$), which seems to lie in an ``HI window''
%out of the Galactic plane (Snowden et al. 1994).

For the remaining four pulsars (with $|b|<7^\circ$), the 21 cm emission
maps do not give meaningful estimates for $N_{HI}$, since much of the
emission is expected to come from behind the pulsar, and much of it may
be optically thick. Therefore, I rely on the rule of thumb mentioned in
the previous paragraph for a crude estimate of $N_{HI}$.

Given the estimates for $d$ and $N_{HI}$, and assuming a fiducial
neutron star radius, $R=10\km$, 
blackbody spectra were folded through the ROSAT position-sensitive
proportional counter (PSPC) response
function (Briel et al. 1994) and HI absorption in order to find the surface 
temperatures, $T_{BB}$, giving the correct count rates, and the 
corresponding ratios of bolometric luminosity to spin-down power, 
$L_{BB}/\dot E_{rot}$. For the two pulsars observed with ROSAT's high-resolution
imager (HRI) rather than the PSPC, the counts were converted to equivalent
PSPC counts by multiplying by a factor of 10, a rough, but reasonable estimate,
given the similarity of the efficiency curves at low energies (Briel et al.
1994). From these results, one obtains the values of $\delta\mu_2$
given in Table 2. Relativistic effects were neglected throughout.

Of course, the assumption of blackbody spectra made above does not strictly
apply. Realistic neutron star atmosphere calculations (Rajagopal \& Romani 1996;
Zavlin, Pavlov, \& Shibanov 1996) yield spectra that depart substantially from a blackbody
with the same effective temperature, in most cases by excess emission in the 
high-energy tail, compensated by a deficit at low energies. Since ROSAT only
sees the high-energy tail, this effect adds to the possible presence of 
additional heat sources in the star and of additional emission processes
around it to make the values obtained here overestimate the true effective 
temperature, luminosity, and $\delta\mu_2$. Thus, the constraints placed on the 
superfluid energy gaps by the values of $\delta\mu_2$ in Table 2 are weaker 
than those which would be obtained with a more complete and realistic model. 

On the other hand, the main difficulty in estimating $\delta\mu_1$ is our
nearly complete ignorance of the true pulsar age, $t$. Varying it
in the range $0<t<t_s$ makes $\delta\mu_1$ sweep from zero to infinity.
Thus, as a very rough approximation, I took $t=t_s/2$, unless this 
turned out larger than the age of the Galactic disk, $\sim 9\Gyr$ 
(Liebert, Dahn, \& Monet 1988, 1989), which is the case for pulsars
J2322+2057 and J2019+2425. For these, I took $t=9\Gyr$. 

Better estimates
of $t$ should soon be available from the luminosities of some pulsars'
white dwarf companions. At present, such estimates are uncertain and 
controversial (e.g., Lorimer et al. 1995; Alberts et al. 1996), and have 
therefore not been taken into account in this work.
%there were strong reasons to disbelieve this estimate. Such a reason 
%existed for PSR J1012+5307, for which the cooling age of the white
%dwarf companion, $t=0.3\Gyr$ (Lorimer et al. 1995), was used instead of 
%$t_s/2=3.5\Gyr$. Similarly, the white dwarf companion of PSR 0751+1807
%implies an age $t\geq 7.9\Gyr$ (Lundgren et al. 1996a), very close to 
%$t_s=8.2\Gyr$, so the lower
%limit on $t$ was adopted as an estimate for the age. Finally, for 
%PSR J2322+2057 and PSR J2019+2425, $t_s/2$ is longer than the
%age of the Galactic disk, $\sim 9\Gyr$ (Liebert et al. 1988, 1989?? CHECK!!), 
%which was adopted as the age for the former, while for the latter I took
%$t=8\Gyr$, the upper limit on its age from observations of its white dwarf 
%companion (Lundgren et al. 1996b). Given $t_s$, $t$, and the pulse period $P$,
%it is straightforward to calculate $\delta\mu_1$.
In addition, more accurate estimates of $t$ will be hampered by the complicated evolutionary
history of ms pulsars (e.g., Bhattacharya \& van den Heuvel 1991). It is
commonly believed that ms pulsars have been spun up by accretion from their
binary companions. In this process, if the temperature does not increase 
enough to allow reactions to restore equilibrium, the core material will 
attain a chemical imbalance opposite in sign to that resulting from spin-down,
which will grow up to the threshold value, $-\delta\mu_{thr}$. Once
mass transfer stops, the star starts spinning down. Now, this spin-down must
first reduce the chemical imbalance to zero before it can start building it up
with the opposite sign. Thus, from the end of the accretion 
phase until reactions resume, the star must lose twice as much rotational 
energy as would be the case if it had started in chemical equilibrium.

\begin{figure}
\plotone{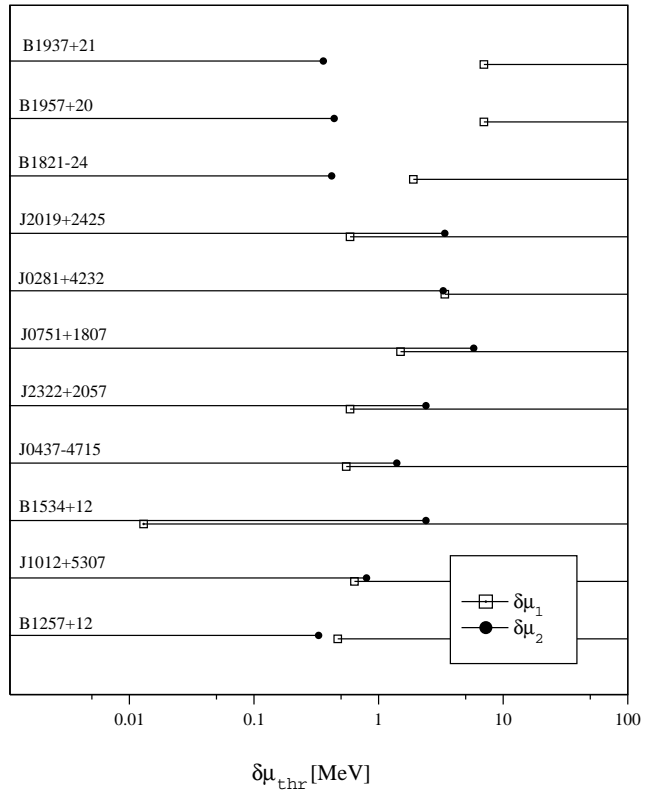}
\caption{Constraints on the threshold chemical potential difference
$\delta\mu_{thr}$ from preliminary analysis of ms pulsar data. For
each pulsar, $\delta\mu_{thr}>\delta\mu_1$ or $\delta\mu_{thr}<\delta\mu_2$.}
\end{figure}

Figure 2 shows the allowed regions $\delta\mu_{thr}>\delta\mu_1$ and 
$\delta\mu_{thr}<\delta\mu_2$ as black horizontal lines, leaving the
forbidden regions (if any) as blanks in between. Though individual
values of $\delta\mu_1$ and $\delta\mu_2$ can hardly be trusted, this 
Figure shows that the current data are in fact starting to probe the 
interesting range $0.1\MeV\lesssim\delta\mu_{thr}\lesssim 1\MeV$. In 
particular, if energy gaps far above $1\MeV$ are considered to be excluded
by theoretical considerations, then the combination of the 3 very fast 
pulsars, B1937+21, B1957+20, and B1821-24, seems to give a fairly safe 
constraint $\delta\mu_{thr}\lesssim 0.4 \MeV$ (unless they are all much younger 
or intrinsically much brighter than assumed here).

\section{Discussion}

The results of the previous section, though far from conclusive in their
present form, show that it might be possible in the near future to use
the present scheme to put interesting constraints on the superfluid
energy gaps in neutron star cores, assuming that all the assumptions made
in the model are correct.

Substantial additional progress can be made along several lines:

1) The theoretical predictions will be made more precise by considering
state-of-the-art neutron star models, their actual contraction due to
spin-down, all particle species present, possibly relevant neutrino 
cooling mechanisms, and the effects of general relativity. Time dependence
is likely to be irrelevant, and a steady-state model should be sufficient.

2) A detailed study of the interaction between the chemistry of the core
and the crust along the lines discussed at the end of \S 2 should definitely
be done. This would show whether reactions in the crust might provide a shortcut
that prevents the growth of a substantial chemical imbalance in the core.
 
3) Deeper x-ray observations (with higher sensitivity and/or longer 
integration time) with better spectral resolution should allow to eliminate 
much of the 
non-thermal contribution to the flux, setting tighter limits on the thermal 
portion and, perhaps, on the amount of absorbing neutral hydrogen along the
line of sight.

4) The improved spectra should be fitted by realistic theoretical atmosphere 
models (Zavlin et al. 1996), rather than
assuming blackbody spectra and constraining them with the total count rates.

5) Measurements of the absorption of pulsar radio flux in the $21\cm$ line 
of neutral hydrogen could improve the estimates of the absorbing column density.

6) An independent limit on the thermal flux might be obtained through optical
observations of the Ray\-leigh-Jeans tail. The predicted fluxes are extremely
low, and in many cases the neutron star may be fainter than its white dwarf 
companion, even in the $B$ and $U$ bands. However, single millisecond
pulsars may be detectable with next-generation telescopes with long enough
integration.

7) Improvements of white dwarf cooling calculations may provide more secure
age determinations for ms pulsar-white dwarf binary systems. This
would help in estimating the initial rotational energy of the pulsar, and 
therefore its expected chemical imbalance, $\delta\mu_1$. 

\acknowledgments

This study has benefitted from information exchange with many colleagues,
including A. Alpar, W. Becker, L. Bildsten, J. Halpern, K. Iida, M. Kiwi, 
P. Krastev, 
J. Miralda-Escud\'e, D. Page, G. Pavlov, M. Ruderman, J. Sauls, and V. Zavlin.
I also thank J. V\'eliz for preparing the figures, and G. Reineking for help
with the tables. This work was financially supported 
by FONDECYT (Chile) grant 1961134 and by a grant from Fundaci\'on Andes' 
``Program for the Insertion of Chilean Scientists.''

\newpage

%\figcaption{Schematic representation of the condition for the occurrence
%of neutron beta decays (direct Urca). 
%(a) $\delta\mu<\delta\mu_{thr}=\Delta_n+\Delta_p$: Reactions not allowed. 
%(b) $\delta\mu>\delta\mu_{thr}=\Delta_n+\Delta_p$: Reactions allowed.}

%\figcaption{Constraints on the threshold chemical potential difference
%$\delta\mu_{thr}$ from preliminary analysis of ms pulsar data. For
%each pulsar, $\delta\mu_{thr}>\delta\mu_1$ or $\delta\mu_{thr}<\delta\mu_2$.}

\begin{deluxetable}{llclrlrlc}
%\tablenum{1}
\tablecolumns{9}
\tablewidth{0pt}
\tablecaption{Upper bounds on the blackbody emission of ms pulsars 
observed by ROSAT}
\tablehead{
\colhead{Name}       &  \colhead{$d$}          &
\colhead{$\log \dot E_{rot}$}          &  \colhead{$R_{PSPC}$}          &
\colhead{$b$}       &  \colhead{$N_{HI}$}          &
\colhead{$kT_{BB}$}       &  \colhead{$L_{BB}/\dot E_{rot}$}          &
\colhead{References}     \\
\colhead{}       &  \colhead{[kpc]}          &
\colhead{[ergs s$^{-1}$]}       &  \colhead{[$10^{-3}{\rm s}^{-1}$]}          &
\colhead{[$^\circ$]}       &  \colhead{[$10^{20}{\rm cm}^{-2}$]}          &
\colhead{[eV]}       &  \colhead{[$10^{-3}$]}          &
\colhead{}}
\startdata
B1257+12   &      ~~0.6  &  33.74  & ~$\leq$ 0.31      &  75 & ~~~~~2.7  & $\leq$  23~  & ~~$\leq$ 0.65  & 1, 2, 3 \nl     
J1012+5307 &      ~~0.5  &  33.50  & ~$\leq$ 5.5       &  51 & ~~~~~0.75 & $\leq$  25~  & ~~$\leq$ 1.6  & 4, 5, 6 \nl    
B1534+12   &      ~~0.7  &  33.28  & ~$\leq$ 1.0       &  48 & ~~~~~3.6  & $\leq$  29~  & ~~$\leq$ 4.7  & 1, 2, 3 \nl    
J0437-4715 &      ~~0.14 &  33.61  &     ~~200.        & -42 & ~~~~~1.9  & $\leq$  33~  & ~~$\leq$ 3.7  & 1, 2, 7, 8 \nl     
J2322+2057 &      ~~0.8  &  33.15  & ~$\leq$ 0.2       & -37 & ~~~~~4.7  & $\leq$  27~  & ~~$\leq$ 4.8  & 1, 2, 3 \nl    
J0751+1807 &      ~~2    &  33.80  &    ~ ~~~3.6       &  21 & ~~~~~4    & $\leq$  49~  & ~$\leq$ 12.   & 9, 10 \nl
J0281+4232 &      ~~5.7  &  35.40  &    ~ ~(21)         & -18 & ~~~~~5    & $\leq$ 107~  & ~~$\leq$ 6.6 & 11, 12 \nl
           &             &         &          &       &      &      &          \nl
J2019+2425 &      ~~0.9  &  33.12  & ~$\leq$ 0.28      & -6.6 & ~~~~~5.3 & $\leq$  29~  & ~~$\leq$ 6.8  & 1, 2 \nl   
B1821-24   &      ~~5.1  &  36.33  &     ~~~10.3       & -5.6 & ~~~~37   & $\leq$ 109~  & ~~$\leq$ 0.84 & 1, 2 \nl    
B1957+20   &      ~~1.5  &  35.06  &     ~~~~2.0       & -4.7 & ~~~~~9   & $\leq$  53~  & ~~$\leq$ 0.87 & 1, 2 \nl    
B1937+21   &      ~~3.6  &  36.04  & ($\leq$ 6.6)       &  2.3 & ~~~~22   & $\leq$  89~  & ~~$\leq$ 0.73 & 2, 12 \nl
\enddata
\tablecomments{Columns: 1. Pulsar name. 2. Adopted distance. 
3. Logarithm of the spin-down power, assuming a moment of inertia
$I=10^{45}{\rm g\,cm}^2$. 4. ROSAT PSPC count rate (numbers in parenthesis
were inferred from HRI observations; see text). 5. Galactic latitude. 
6. Adopted column density of neutral hydrogen along the line of sight to the 
pulsar. 7. Highest blackbody temperature of
a spherical object of 10 km radius at the adopted distance compatible with
the detected count rates (or upper limits) and the adopted HI column density. 
8. Ratio of the blackbody luminosity to the spin-down energy. 9. References
for the numbers in columns 1-6: (1) Danner et al. (1994); 
(2) Taylor et al. (1993); (3) Burstein \& Heiles (1978); 
(4) Nicastro et al. (1995); (5) Lorimer et al. (1995); (6) Halpern (1996); 
(7) Bell et al. (1995); (8) Heiles \& Cleary (1979); (9) Lundgren et al. 
(1995); (10) Becker et al. (1996); (11) Navarro et al. (1995);
(12) Verbunt et al. (1996).}
\end{deluxetable}

%\newpage

\begin{deluxetable}{lllllc}
%\tablenum{2}
\tablecolumns{6}
\tablewidth{0pt}
\tablecaption{Constraints on $\delta\mu_{thr}$ from ms pulsars observed by
       ROSAT}
%\begin{center}
\tablehead{
\colhead{Name}  &  \colhead{$P$}  &  
\colhead{$t_s$}  &  \colhead{$\delta\mu_1$}  &  
\colhead{$\delta\mu_2$}  &  \colhead{References} \\
\colhead{}  &  \colhead{[ms]}  &  
\colhead{[Gyr]}  &  \colhead{[MeV]}  &  
\colhead{[MeV]}  &  \colhead{}}
\startdata
B1257+12    & ~~6.2 & ~~3.2 & ~~0.47  & ~~0.33 & 1 \nl
J1012+5307  & ~~5.3 & ~~7   & ~~0.64 & ~~0.8 & 4, 5  \nl
B1534+12    & ~37   & ~~0.2 & ~~0.013 & ~~2.4 & 1  \nl
J0437-4715  & ~~5.7 & ~~5   & ~~0.55  & ~~1.4 & 7  \nl
J2322+2057  & ~~4.8 & ~21   & ~~0.59  & ~~2.4 & 1 \nl
J0751+1807  & ~~3.5 & ~~8.2 & ~~1.5    & ~~5.8 & 9 \nl
J0281+4232  & ~~2.3 & ~~0.5 & ~~3.4   & ~~3.3 & 11 \nl
            &      &      &       &       \nl
J2019+2425  & ~~3.9 & ~27   & ~~0.59  & ~~3.4  & 1 \nl
B1821-24    & ~~3.1 & ~~0.03& ~~1.9   & ~~0.42 & 1 \nl
B1957+20    & ~~1.6 & ~~2.1 & ~~7.0   & ~~0.44 & 1 \nl
B1937+21    & ~~1.6 & ~~0.2 & ~~7.0   & ~~0.36 & 2 \nl
\enddata
%\end{center}
\tablecomments{Columns: 1. Pulsar name. 2. Radio pulse period. 3. 
Characteristic (spin-down) age, $t_s=P/(2\dot P)$. 
%4. Adopted true age,
%in cases where it is not taken as $t=t_s/2$ (see discussion in the text). 
4. $\delta\mu_1$ (defined in the text), as calculated from the data in 
columns 2 and 3. 5. $\delta\mu_2$ (defined in the text), as calculated from the 
data in Table 1. 
6. References for the numbers in columns 2 and 3, as listed below Table 1.}
\end{deluxetable}

\end{document}